\documentstyle[12pt]{article}

\begin{document}

\begin{flushright}
math-ph/0012045
\end{flushright}

\vspace{5mm}

\begin{center}
{\Large \sf Self-dual Chern-Simons Vortices on Riemann Surfaces}\\[10mm]
{\sf Seongtag Kim}\footnote{Electronic mail: stkim$@$skku.ac.kr} \\
{\it Department of Mathematics and Institute of Basic Science,
Sungkyunkwan University,\\
Suwon 440-746, Korea}\\[2mm]
{\sf Yoonbai Kim}\footnote{Electronic mail: yoonbai$@$skku.ac.kr} \\
{\it BK21 Physics Research Division and Institute of Basic Science,
Sungkyunkwan University,\\
Suwon 440-746, Korea}
\end{center}

\vspace{5mm}

\begin{abstract}
We study self-dual multi-vortex solutions of Chern-Simons Higgs
theory in a background curved spacetime. The existence and
decaying property of a solution are demonstrated.
\end{abstract}

\newpage

\def\bea{\begin{eqnarray}}
\def\eea{\end{eqnarray}}
\def\be{\begin{equation}}
\def\ee{\end{equation}}

\setcounter{equation}{0}
\section{\large{\sf INTRODUCTION}}

Chern-Simons gauge theories have provide intriguing questions and
answers to various subjects of both physics and mathematics. One
of interdisciplinary topics attracted attention is so-called
self-dual Chern-Simons solitons~\cite{HKP,JLW,Dun}. A natural
extension is to include gravity which can be
background~\cite{Sch,CK} or dynamical~\cite{Val,CL}. Once the
Bogomolnyi-type bound is obtained and detailed mathematical
properties of those self-dual vortices are studied in the
Chern-Simons Higgs model in the presence of background gravity, it
would be helpful to address related physics-wise problems involving
condensed matter systems, e.g., quantum Hall effects,
supergravity, Lorentz-symmetry breaking due to parity-violating
term, existence of time-like closed curve around gravitating
spinning strings, cosmological implication of cosmic strings, and
even cosmological constant problem.

Because it is applicable to diverse fields, mathematical study of
self-dual Chern-Simons solitons is going on. The existence of a
topological multi-vortex solution of relativistic Chern-Simons
Higgs theory in flat $R^{2}$ is shown  by Wang~\cite{Wan}. In the
same setting, rotationally-symmetric nontopological solitons and
vortices were proven to exist by Spruck and Yang~\cite{SY}. Yang
also proved the existence of a topological self-dual multi-vortex
solution when the gauge symmetry is extended to
non-Abelian~\cite{Yan2}. When the topological vortices or
nontopological solitons are generated in condensed matter systems
or in the early universe, they are likely to form a lattice
structure or a network. In such sense important works have been
done on torus~\cite{CY,Tar,DJLW1,NT} or on standard
sphere~\cite{Sch,DJLW2}. Condensed matter experiments are usually
performed by turning on constant external electric or magnetic
field. In relation to this, Chae {\it et al.} demonstrated the
existence of soliton solutions of self-dual Chern-Simons Higgs
model coupled to an external background charge
density~\cite{Chae}. Another study to have cosmological
implication was done by Choe with nontopological soliton solutions
under decaying metric~\cite{Choe}.

Now let us take into account  curved spacetime geometry of a
straight string in the early universe. Then,  extremely-small core
region of the string is curved by matter fields, and the
intermediate region is slightly-curved or locally-flat because of
no graviton to the transverse directions. However, the asymptote
of the global universe is known to be flat. All of such geometry
should be dynamically determined by examining Einstein equations
in exact sense, but it is practically too difficult to do with
mathematical rigor. A meaningful starting point is to assume a
physically-allowable set of background metrics and to study
possible string configurations. In this paper, we study
Chern-Simons Higgs theory on a uniformly Euclidean metric, which
is not necessarily radial. A spatial metric 
$\gamma_{ij} = b(x,y) \delta_{ij}$
is called  uniformly Euclidean metric if  there exist positive
constants $a_1$ and $a_2$ with $a_1 \le b(x,y) \le a_2$. We show
the existence of a self-dual topological multi-vortex solution and
the fast decay property of a solution at infinity. The
mathematical conditions we bring up are relevant to the physical
situation discussed in the above, e.g., the gravity is not far
from that of the flat case at the end of universe.

A brief outline of the paper is in order. In section 2, under the
most general static metric, we shall derive the Bogomolnyi type
bound of the Chern-Simons Higgs theory in background gravity. In
section 3, we present the existence and asymptotic behavior of a
solution of the self-dual Chern-Simons vortices. Conclusions with
some discussions about our results are presented in section 4.

\setcounter{equation}{0}
\section{\large{\sf BOGOMOLNYI BOUND OF CHERN-SIMONS HIGGS THEORY IN
BACKGROUND GRAVITY}}

In this section we recapitulate derivation of so-called Bogomolnyi bound of
the
Chern-Simons Higgs theory coupled to background gravity by assuming the
general static metric
\begin{equation}
ds^{2}=N^{2}(x^{k})dt^{2}-\gamma_{ij}(x^{k})dx^{i}dx^{j} ~~~
(i,j,k,... =1,2),
\end{equation}
where the metric of two-dimensional spatial hypersurface can
always be  diagonalized by a conformal gauge
$\gamma_{ij}=\delta_{ij}b(x^{k})$. Later we shall show that the
Bogomolnyi bound is attained only when the lapse function
$N(x^{i})$ is constant, i.e., $N(x^{k})=1$ after a rescaling of
time coordinate $t$.

The Chern-Simons Higgs theory is described by the action
\begin{equation}
S=\int d^{3}x\sqrt{g}\left[
\frac{\kappa}{2}\frac{\epsilon^{\mu\nu\rho}}{\sqrt{g}}
A_{\mu}\partial_{\nu}A_{\rho} +\frac{1}{2}g^{\mu\nu}\overline{D_{\mu}\phi}
D_{\nu}\phi-V(|\phi|) \right],
\end{equation}
where $\phi=e^{i\Theta}|\phi|$ is a complex scalar field,
$A_{\mu}$ a U(1) gauge field, and
$D_{\mu}=\partial_{\mu}-ieA_{\mu}$ is gauge-covariant but not
covariant under general coordinate transformation. Since the
Bogomolnyi limit is our interest, the form of the scalar potential
$V(|\phi|)$ is taken to be
\begin{equation}\label{pot}
V(|\phi|)=\frac{e^{4}}{8\kappa^{2}}|\phi|^{2}(|\phi|^{2}-v^{2})^{2}.
\end{equation}
From here on all the metric components and fields are assumed to be static
because the self-dual solitons of our interests are static objects.

Symmetric energy-momentum tensor is
\begin{equation}
T_{\mu\nu}=\frac{1}{2}(\overline{D_{\mu}\phi}D_{\nu}\phi+
\overline{D_{\nu}\phi}D_{\mu}\phi)-g_{\mu\nu}\left[\frac{1}{2}g^{\rho\sigma}
\overline{D_{\rho}\phi}D_{\sigma}\phi-V(|\phi|)\right].
\end{equation}
A physically-meaningful derivation of the Bogomolnyi bound is to
investigate vanishing of stress components of the energy-momentum
tensor.  Since the lapse function $N(x^i)$ disappears in every stress
component by the help of Gauss' law $\kappa
NB=e^{2}A_{0}|\phi|^{2}$, an appropriate rearrangement of them
gives
\begin{eqnarray}
T^{ij}&=&\frac{1}{2}\gamma^{ij}\left[\frac{\kappa^{2}}{2e^{2}}
\frac{B^{2}}{|\phi|^{2}}-V(|\phi|)\right]-\frac{1}{2}(\gamma^{ij}\gamma^{kl}
-\gamma^{ik}\gamma^{jl}-\gamma^{il}\gamma^{jk})\overline{D_{k}\phi}
D_{l}\phi \\
&=&\frac{\kappa^{2}}{2e^{2}}\frac{\gamma^{ij}}{|\phi|^{2}}
\left[B-\frac{e^{3}}{2\kappa^{2}}|\phi|^{2}(|\phi|^{2}-v^{2})\right]
\left[B+\frac{e^{3}}{2\kappa^{2}}|\phi|^{2}(|\phi|^{2}-v^{2})\right]\nonumber\\
&&+\frac{1}{8}\left\{
\left[\left(\overline{D^{i}\phi\mp i\frac{\epsilon^{ik}}{\sqrt{\gamma}}
\gamma_{kl}D^{l} \phi}\right)
\left(D^{j}\phi\pm i\frac{\epsilon^{jm}}{\sqrt{\gamma}}
\gamma_{mn}D^{n} \phi\right) \right. \right.\nonumber\\
&&\hspace{14mm}
\left. +\left(\overline{D^{j}\phi\pm i\frac{\epsilon^{jk}}{\sqrt{\gamma}}
\gamma_{kl}D^{l} \phi}\right)
\left(D^{i}\phi\mp i\frac{\epsilon^{im}}{\sqrt{\gamma}}
\gamma_{mn}D^{n} \phi\right) \right] \nonumber\\
&&\hspace{6mm}
+\left[\left(\overline{D^{i}\phi\pm i\frac{\epsilon^{ik}}{\sqrt{\gamma}}
\gamma_{kl}D^{l} \phi}\right)
\left(D^{j}\phi\mp i\frac{\epsilon^{jm}}{\sqrt{\gamma}}
\gamma_{mn}D^{n} \phi\right) \right.\nonumber\\
&&\hspace{14mm}\left.\left. + \left.
\right.\left(\overline{D^{j}\phi\mp
i\frac{\epsilon^{jk}}{\sqrt{\gamma}} \gamma_{kl}D^{l} \phi}\right)
\left(D^{i}\phi\pm i\frac{\epsilon^{im}}{\sqrt{\gamma}}
\gamma_{mn}D^{n} \phi\right) \right]\right\}  \label{tij},
\end{eqnarray}
where $\gamma^{ij}$ is inverse of the $\gamma_{ij}$, $\sqrt{\gamma}=\sqrt{\det
\gamma_{ij}}$, and the magnetic field is defined by
$B=-\frac{\epsilon^{ij}}{\sqrt{\gamma}}\partial_{i}A_{j}$.

We read the first-order Bogomolnyi equations from Eq.~(\ref{tij})
\begin{equation}\label{bog1}
B=\mp\frac{e^{3}}{2\kappa^{2}}|\phi|^{2}(|\phi|^{2}-v^{2}),
\end{equation}
\begin{equation}\label{bog2}
D_{i}\phi\mp i\sqrt{\gamma}\epsilon_{ij}\gamma^{jk}D_{k}\phi=0.
\end{equation}
The second equation (\ref{bog2}) expresses the spatial components
of the gauge field  $A_i$ in terms of the scalar field, i.e.,
$eA_{i}=\partial_{i}\Theta
\mp\sqrt{\gamma}\epsilon_{ij}\gamma^{jk}\partial_{k}\ln |\phi|$.
Substituting it into the first Bogomolnyi
equation (\ref{bog1}) together with the conformal gauge, we have
\begin{equation}\label{boge}
\partial^{2}\ln
|\phi|=\frac{e^{4}}{2\kappa^{2}}b|\phi|^{2}(|\phi|^{2}-v^{2})
\mp\epsilon^{ij}\partial_{i}\partial_{j}\Theta ,
\end{equation}
where Dirac-delta function like contribution of the scalar phase 
$\Theta$ comes from
multi-valued function such as $\Theta=\sum_{k=1}^{n}\tan^{-1}
\frac{x^{2}-x^{2}_{k}}{x^{1}-x^{1}_{k}}$.

Let us check a consistency condition that whether or not the Bogomolnyi
equations (\ref{bog1})$\sim$(\ref{bog2}) reproduce second-order
Euler-Lagrange equations. Since we used the Gauss' law, let us consider
scalar field equation;
\begin{equation}\label{sca}
\frac{1}{\sqrt{g}}D_{\mu}(\sqrt{g}g^{\mu\nu}D_{\nu}\phi)=-\frac{\phi}{|\phi|}
\frac{dV}{d|\phi|}.
\end{equation}
For a static configuration, insertion of the Bogomolnyi equations
(\ref{bog1}), (\ref{bog2}), (\ref{bog2}) into the scalar equation
(\ref{sca}) leads to
\begin{equation}
\frac{1}{N}\gamma^{ij}\partial_{i}N\partial_{j}|\phi|=0 .
\end{equation}
As it is well-known, for every configuration of the self-dual solitons,
derivative of the scalar amplitude vanishes nowhere, and both derivatives,
$\partial_{i}N$ and $\partial_{j}|\phi|$, are not perpendicular each other.
Then, the lapse function $N$ should be
a constant which we set to be one by a rescaling of time variable, i.e.,
$dt\rightarrow dt/N$. Note that the spatial components of the gauge-field
equation are automatically reproduced for $N=1$ without giving any additional
constraint.

Now that we have the condition $N=1$, the derivation of the Bogomolnyi bound
reduces to the original one by Schiff~\cite{Sch}. The energy is exactly
proportional to the magnetic flux $\Phi=\int d^{2}x\sqrt{\gamma}B$ as follows
\begin{eqnarray}
E&=&\int d^{2}x\sqrt{\gamma}
\left[\frac{\kappa^{2}}{2e^{2}}\frac{B^{2}}{|\phi|^{2}}
+\frac{1}{2}\gamma^{ij}\overline{D_{i}\phi}D_{j}\phi+V(|\phi|) \right]
\nonumber\\
&=&\int d^{2}x\sqrt{\gamma}
\left\{\frac{\kappa^{2}}{2e^{2}}\frac{1}{|\phi|^{2}}
\left[B\pm\frac{e^{3}}{2\kappa^{2}}|\phi|^{2}(|\phi|^{2}-v^{2})\right]^{2}
\right.  \nonumber\\
&&\hspace{22mm}+\frac{1}{4}\gamma^{ij}
(\overline{D_{i}\phi\mp i\sqrt{\gamma}\epsilon_{ik}\gamma^{kl}D_{l}\phi})
(D_{j}\phi\pm i\sqrt{\gamma}\epsilon_{jm}\gamma^{mn}D_{n}\phi)
\nonumber\\
&&\hspace{22mm}\left.\pm\frac{ev^{2}}{2}B\pm\frac{i}{4}\frac{1}{\sqrt{\gamma}}
\partial_{i}\left[\epsilon^{ij}(\bar{\phi}D_{j}\phi-\overline{D_{j}\phi}\phi)
\right]\right\}
\label{esh}\\
&\ge&\left|\frac{ev^{2}}{2}\Phi \right|.
\end{eqnarray}
The first and second lines of Eq.~(\ref{esh}) vanish by substituting the
Bogomolnyi equations (\ref{bog1})$\sim$(\ref{bog2}), and the last 
total-divergence term in the third line of Eq.~(\ref{esh}) does not 
contribute to the energy since U(1) current decays rapidly at spatial asymptote.

We read possible boundary conditions of the scalar amplitude at
spatial infinity from the scalar potential (\ref{pot}), that is,
$\lim_{|x|\rightarrow\infty}|\phi|\rightarrow 0~{\rm or}~v$. The
former is a nontopological soliton or vortex, and the latter
a topological vortex. All of them carry the magnetic flux $\Phi$ (or
equivalently U(1) charge $Q=e\int
d^{2}x\sqrt{\gamma}A_{0}|\phi|^{2}$ related by the Gauss' law),
and spin
\begin{eqnarray}
J&\equiv&\int
d^{2}x\sqrt{\gamma}\sqrt{\gamma}\epsilon_{ij}x^{i}T^{j}_{\;0} \\
&=&\int
d^{2}x\sqrt{\gamma}\frac{1}{2}\sqrt{\gamma}\epsilon_{ij}x^{i}
(\overline{D^{j}\phi}D_{0}\phi+\overline{D_{0}\phi}D^{j}\phi)\\
&=&\frac{e^{2}}{8\kappa}\int d^{2}x\sqrt{\gamma}x^{i}\partial_{i}
(|\phi|^{2}-v^{2})^{2},
\end{eqnarray}
which distinguishes the Chern-Simons solitons from the solitons in Abelian
Higgs model.

\setcounter{equation}{0}
\section{\large{\sf EXISTENCE OF A SOLUTION}}
Throughout this section, we denote that $(M,\gamma)$ is a
two-dimensional complete  Riemann surface which is diffeomorphic
to $R^2$ with the metric $\gamma_{ij}= b(x,y) \delta_{ij}$. We
assume that there exist positive constants $a_1$ and $a_2$ with
$a_1 \le b(x,y)\le a_2$ for all $z=(x,y)\in R^2 $ ($x^1=x$ and
$x^2=y$ from here on). Let $\Delta ={1 \over {\sqrt
{det(\gamma_{ij})}}} ({{\partial^2}\over{\partial x^2}}
+{{\partial^2}\over{\partial y^2}})$  ($\Delta_0
={{\partial^2}\over{\partial  x^2}} +{{\partial^2}\over{\partial
y^2}}$), $|\nabla u| (  |\nabla u|_{E})$ and  $\delta $
($\delta_{E} $) is the Laplacian, the norm of the gradient and
Dirac-delta function with respect to the metric $\gamma_{ij}$
(Euclidean metric). We denote $dz=dx dy$, $dV_{\gamma}=b(x,y)dz$
and $ H_1^2 $ be the Sobolev space, which is the completion of
$C_c^{\infty}(M)$ with respect to  the norm $||w||=\big( \int_{M}
|\nabla w|^2 +w^2 \ dV_{\gamma})^{1\over 2}$. \vskip 0.3 true cm

 In this section, we show   the following Theorem.
\vskip 0.2 true cm \noindent
 \bf Theorem 1:\rm \quad
There exists a solution for the following self-dual Chern-Simons
vortex equation on $(M,\gamma)$
\begin{equation}\label{mathe}
 \Delta w= e^w (e^w-1) +  4 \pi \sum _{k=1}^{n}\delta(z-z_k)
 \label{c1}
\end{equation}
\noindent with the boundary condition $ \lim_{|z|\to \infty} w=0 $.  Moreover, 
$w$ satisfies $ -a e^{-b |x|} \le w(x)<0$ at infinity for some positive
constants $a$ and $b$.

\vskip 0.2 true cm

The above equation (\ref{mathe}) comes from Eq.~(\ref{boge}) by rescaling
the scalar field $|\phi|=ve^w$ and the spatial coordinates
$x^{i}\rightarrow\frac{\kappa}{e^{2}v^{2}}x^{i}$.

\vskip 0.2 true cm

When there is no vortex, there is no $H_{1}^2$ solution for the
following  equation (\ref{d1}), \bea\label{d1} \Delta w= e^w
(e^w-1), \eea \noindent because \bea\label{d2} \int_{\Omega} e^w
(e^w-1) w dV_{\gamma} = -\int_{\Omega} |\nabla w|^2 dV_{\gamma} +
\int_{\partial\Omega} w {{\partial w}\over {\partial \eta}} dS
\eea
 for a sufficiently large smooth domain  $\Omega $ (see Ref. \cite{Sch} ).
Applying the same method to the domain outside of vortices,
 we see that any $H_1^2$ solution $w$ of Eq.~(\ref{c1}) satisfies
 $w\le 0$.
\vskip 0.2 true cm
 \it  Proof of Theorem 1 \rm: \quad
  To show the existence of a solution, we follow \cite{Wan}.
  Take $u_0$ be  \be u_0=
-\sum_{k=1}^{n} ln(1+\mu |z-z_k|^{-2}), \label{c2} \ee then
 \bea \Delta_0u_0=-4 \sum _{k=1}^{n}
{{\mu}\over { (\mu+|z-z_k|^2)^2}}+4 \pi \sum
_{k=1}^{n}\delta_E(z-z_k). \label {c3} \eea

\noindent  Note that for any given smooth function $f(z)$, \bea
\int_{M} \Delta ln|z-z_k|^2 f(z) dV_{\gamma} &=& \int_{M} {1
\over{b(x,y)}} \Delta_0 ln|z-z_k|^2 f(z) b(x,y) dx dy  \nonumber \\
&=& \int_{R^2} \Delta_0 ln|z-z_k|^2 f(z)  dx dy \nonumber \\ &=&
4\pi f(z_k). \eea \noindent
 Therefore, $\Delta ln |z-z_k|^2= 4 \pi
\delta(z-z_k)$ and $\Delta_E ln |z-z_k|^2=4 \pi \delta_E(z-z_k)$.
Define $h_0$ , $h$ and $B$ as the followings,
 $$ h_0=4 \sum_{k=1}^{n} {{\mu}\over {
(\mu+|z-z_k|^2)^2}}, \ h=h_0/b, $$ \noindent  and
 \bea B=e^{u_0}=\prod_{k=1}^{n} {{
|z-z_k|^2}\over{\mu+|z-z_k|^2}}.\eea \noindent
 Take $w=u_0+u$, then Eq.~(\ref{c1})
turns out to be
\begin{equation}
 \Delta u = B e^u (B e^u-1) +  h.   \label{c9}
\end{equation}
\noindent
 A critical point of  functional $E$ defined on $H_1^2$ is a solution
of Eq.~(\ref{c9}) where \bea E(u)=
  \int_{M} |\nabla u|^2 +(
B e^u-1)^2+2 h u \, dV_{\gamma} . \label{c10} \eea \noindent By
the basic inequality,  \bea (e^u-1)^2 \ge { |u|^2 \over{ (1+
|u|)^2}}, \label{c18}\eea
 \noindent
  and \bea
 \int_{M} 2 hu \, dV_{\gamma} &=& \int_{R^2} 2 h_0 \,u
\,dz \nonumber \\ &\le& 2 \left(\int_{R^2} h_0^2 dz
\right)^{1\over 2} \left(\int_{R^2} u^2 dz \right)^{1\over 2} .
  \eea
Note that there exist constants $c_1$ and $c_2$ such that
 \bea 2 \left(\int_{R^2} h_0^2 dz\right)^{1 \over 2} \le { {c_1} \over {\sqrt{\mu}}}
 \eea
 \noindent and \bea
\int_{M}(B-1)^2 dV_{\gamma} \le c_2 . \label{c316} \eea \noindent
Note that Eq.~(\ref{c316}) holds when $dV_{\gamma} <c
r^{3-\epsilon} dr$ for any positive constant $c$ and any positive
small constant $\epsilon$. The second term of Eq.~(\ref{c10}) can
be estimated as
 \bea
\int_{M} (B e^u-1)^2 dV_{\gamma} &\ge& {1 \over 2} \int_{M}
B^2(e^u -1)^2-(B-1)^2 dV_{\gamma} .\label{c12}\eea \noindent
 Let us define
$\Omega_1=\{x \in M| B^2(x) \le 1/2\} $ and $|\Omega_1|$ be the
area of $\Omega_1$. The finiteness of $|\Omega_1|$ implies
 \bea \int_{\Omega_1} \left(B^2- {1 \over 2} \right) { {|u|^2} \over{ (1+ |u|)^2}} dV_{\gamma} &\ge&
\int_{\Omega_1} - {1 \over 2} { {|u|^2} \over{ (1+ |u|)^2}}
dV_{\gamma} \nonumber
\\ &\ge& - {1 \over 2} |\Omega_1| . \label{c20} \eea
\noindent
 From Eqs.~(\ref{c18}) and (\ref{c20}), there is a constant $c_3$
that
 \bea \int_{M} B^2(e^u-1)^2 dV_{\gamma} &\ge&
  \int_{M} { {B^2|u|^2} \over{ (1+ |u|)^2}} dV_{\gamma} \nonumber \\
  &=&\int_{M-\Omega_1} { {B^2|u|^2} \over{ (1+ |u|)^2}} dV_{\gamma} +
  \int_{\Omega_1} { {B^2|u|^2} \over{ (1+ |u|)^2}} dV_{\gamma} \nonumber \\
 %%%%%%%%%%%%%%%%%%%
 &\ge&
 {1 \over2} \int_{M} { |u|^2 \over{ (1+ |u|)^2}}
 dV_{\gamma} -c_3
  \label{c22}\eea
%%%%%%%%%%%%%%%%%%%
For $f \in H_1^1(R^2)$, $\int_{R^2}f^2 dz \le { 1 \over 4}  \Big(
\int_{R^2} |\nabla f| dz \Big)^2 $. Set $ f=u^2$ and we have \bea
 \int_{R^2} u^4 dz &\le& \left( \int_{R^2} | u \nabla u|
dV_{\gamma}\right)^2
\nonumber   \\
 &\le&  \int_{R^2}  u^2 dz    \int_{R^2}  |\nabla
u|^2 dz , \label{c31}  \eea \noindent and
 \bea
 \left(\int_{R^2} u^2 dz \right)^2 &\le &
  \left[ \int_{R^2}  \left(    {|u| \over { 1+ |u|}} \right) (
1+|u|) |u|   dz \right]^2 \nonumber   \\ &\le&  \int_{R^2} \left(
{ |u| \over { 1+|u|} } \right)^2 dz
 \int_{R^2} u^2 (1+|u|)^2   dz    \nonumber \\
  &\le&  2 \int_{R^2} \left(  {  |u| \over {
1+|u|} } \right)^2 dz
 \int_{R^2} u^2 + u^4   dz  . \label{c34} \eea

\noindent
 Using Eqs.~(\ref{c31}) and (\ref{c34}), we obtain
\bea
 \int_{R^2} u^2 dz  \le
2 \int_{R^2} \left(  {  |u| \over { 1+|u|} } \right)^2 dz \ \left(
1+ \int_{R^2} |\nabla u|^2 dz  \right). \label{c344} \eea
\noindent By the H\"{o}lder inequality and Eq.~(\ref{c344}),  $ u
\in H_1^2(R^2)$ satisfies
 \bea \left(\int_{R^2} u^2 dz \right)^{1 \over 2} \le
\int_{R^2} { |u|^2 \over{ (1+ |u|)^2}} dz + 2 \int_{R^2} |\nabla
u|_{E}^2 dz +2. \label{c27} \eea \noindent From the above, there
exists a constant $c_4$ such that \bea E(u)&\ge& \int_{M}|\nabla
u|^2 dV_{\gamma} +{1 \over 2} \int_{M} { |u|^2 \over{ (1+ |u|)^2}}
dV_{\gamma} \nonumber
\\ &{}& \hbox{\quad} -{c_1 \over {\sqrt{\mu}} } \left[ \int_{R^2} {
|u|^2 \over{ (1+ |u|)^2}} dz + 2 \int_{R^2} |\nabla u|_E^2 dz +2
\right] -c_3-c_2
\\
&\ge& \left(1- { {2c_1} \over {\sqrt{\mu}}} \right)
\int_{M}|\nabla u|^2 dV_{\gamma} + \left( {1 \over2}-{{c_1}\over
{a_1 \sqrt{\mu}}} \right)\int_{M} { |u|^2 \over{ (1+ |u|)^2}}
dV_{\gamma} -c_4, \label{c331}
 \eea
 \noindent
where we used $\int_{M}|\nabla u|^2 dV_{\gamma}= \int_{R^2}|\nabla
u|_E^2 dz .$
 From Eq.~(\ref{c27}) and by taking large $\mu$, there exist
constants $c_5$ and $c_6$ such that \bea E(u)
 \ge c_5
\Big( \int_{M}|\nabla u|^2 + u^2  dV_{\gamma} \Big)^{1/2} -c_6 .
\eea \noindent Therefore $E(u)$ is coercive on $H_1^2$ and
$\inf_{u\in H_1^2}E(u)$ is finite. Moreover, $E(u)$  is weakly
lower semi-continuous functional on $H_1^2$. We take a minimizing
sequence $\{u_n\}$ for $\inf_{u\in H_1^2}E(u)$.  Then $\{u_n\}$ is
bounded on $H_1^2$, which has a subsequence $\{u_{n_k}\}$
converging to $u\in H_1^2$, a minimizer for $\inf_{u\in
H_1^2}E(u)$. By the elliptic regularity, $u$ is smooth. Finally,
$u$ satisfies Eq.~(\ref{c9}). \vskip 0.5 true cm

\vskip 0.3 true cm

Next we study the behavior of solution of Eq.~(\ref{c1}). Since
$\Delta w \le 0$ and $w\le 0$, we have $ -c ||w||_{L^2(B(x,1))}
\le w(x) <0 $
  at  infinity for some positive constant $c$ (see Ref. \cite{Gil}).
Therefore $w$ decays to zero uniformly at infinity.
 For a sufficiently small positive constant $\delta$,
  $ -\delta <w <0$ implies
\bea
  w \Delta_0 w &=& b \ e^w (e^w-1)w \nonumber  \\ &\ge& {w^2 \over 2} .\eea
    By Jaffe and Taubes
\cite{jt} methods (page 83), $ -a e^{-b |x|} \le w(x) <0 $ for
some positive constants $a$ and $b$ at infinity.

Remarks:   Since  $a_2$ does not appear in Eq.~(\ref{c331}), we
can generalize Theorem 1 if the integral value of  Eq.~(\ref{c316})
is bounded. For example, Theorem 1 holds if $c' dr \le
dV_{\gamma}=b(x,y) dz \le c r^{3-\epsilon} dr $ at infinity for
any positive constants $c'$, $c$ and any small positive constant
$\epsilon$.
%%%%%%%%%%%%%%%%%%%%%%%%%%%%%%%%%%%%%%%%%%%%%%%%%%%%%%%%%%%%%%%%%%%%%%%%%%

\setcounter{equation}{0}
\section{\large{\sf CONCLUDING REMARKS}}
We extend the existence and decay property of topological
multi-vortex solutions of Chern-Simons Higgs theory in a general
background curved spacetime, which have been studied on flat space
or on special background metric. Finding the borderline of growth
or decaying condition of the  given metric, which gives Theorem 1,
is an interesting question. Related issues, e.g., the existence of
nontopological solitons and vortices, self-dual topological
vortices in a suitably decaying metric and Chern-Simons solitons
under a dynamical gravity, need further study.

%%%%%%%%%%%%%%%%%%%%%%%%%%%%%%%%%%%%%%%%%%%%%%%%%%%%%%%%%%%%%%%%%%%%%%%%%%
\section*{\large{\sf ACKNOWLEDGMENTS}}
This work is supported by Faculty Research Fund, Sungkyunkwan
Univ. 1999 (S.K.) and No. 2000-1-11200-001-3 from the Basic
Research Program of the Korea Science $\&$ Engineering Foundation
(Y.K.).

\end{document}